\documentclass[11pt]{article}
\usepackage[pdftex,colorlinks=true,urlcolor=black,filecolor=black,linkcolor=black,
            pdftitle={Rotational Dimensional Reduction.},
            pdfauthor={Mark D. Roberts},
            pdfsubject={Rotational Dimensional Reduction},
            pdfkeywords={solar system, dimensional reduction},
            pagebackref,pdfpagemode=None,bookmarksopen=true]{hyperref}
\newcommand{\bc}{\begin{center}}
\newcommand{\ec}{\end{center}}
\newcommand{\be}{\begin{equation}}
\newcommand{\ee}{\end{equation}}
\newcommand{\ber}{\begin{eqnarray}}
\newcommand{\ear}{\end{eqnarray}}

\begin{document}
\title{Rotational Dimensional Reduction.}
\author{
\href{http://www.violinist.com/directory/bio.cfm?member=robemark}
{Mark D. Roberts},\\
}
\date{$1^{st}$ of November 2016}
\maketitle
\begin{abstract}
Rotational energy dissipation in the solar system confines the planets to the ecliptic,
this can be thought of as a dimensional reduction from three dimensions to two.
It is argued that the same mechanism restricts five dimensional matter to four dimensional
spacetime.  The result is sensitive to geometric configuration but not to force law.
Although the mechanism provides a qualitative description it as yet makes no quantitative prediction.
\end{abstract}
\vspace{1cm}
\begin{center}
{\sc keywords:  dimensional reduction,  rotational energy dissipation,  solar system.}
\end{center}
{\tableofcontents}
\newpage
\section{Introduction.}\label{intro}
In the solar systema and similar astrophysical systems roatational dissipation energy
causes most matter to be close to the ecliptic.
It is natural to ask whether such a mechanism occurs in other physical systems.
In particular whether five dimensional matter can be ''matter trapped'' \cite{BBM}\cite{mdr48},
in other words restricted to four dimensional spacetime.
Here it is argued that a rotation ''into'' the fifth dimension can do this,
it is important to note that this rotation is intrinsically five dimensional and not
about some four dimensional brane point,  and so leads to no observable centre of rotation.
All matter is treated as ''test'' matter which does not contribute explicitly to the background
field:  so when dealing with the solar system all the test matter ($\sum_i m_i$) represents
planets and so forth whereas the background field ($GM/r$) determines the dynamics;
similarly for the five dimensional case the test matter can be thought of as objects typical
in the observable universe and the backgound field as the gravitational field associated
with five dimensional de Sitter space.
The conventions used are those of \cite{HE}.
\section{Newtonian three dimensional solar system description.}\label{newtdescription}
Variation of the lagrangian can be expressed as
\begin{equation}
\label{varl}
\delta{\cal L}=\delta E-\lambda\cdot\delta L
\end{equation}
where $\lambda$ is a lagrange multiplier and
\begin{eqnarray}
\label{begineqs}
&&E\equiv KE+PE,\\
&&KE=\frac{1}{2}\sum_i m_i \vec{v}^2_i,~~~
PE=\sum_i\phi_i,\nonumber\\
&&L=\sum_i m_i \vec{r}_i\times\vec{v}_i
=\sum_i m_i\epsilon_{\alpha\beta\delta}r^\beta_i v^\delta_i,
\nonumber
\end{eqnarray}
where $E$ is the energy,  $KE$ is the kinetic energy of the test matter,
$PE$ is the potential energy of the background field,
$L$ is the sum of the angular momenta of the particles
and $i$ labels each particle,  from now on drop it and vector overscore.
Using (\ref{begineqs}) in (\ref{varl}) and assuming that $\delta v$ and $\delta r$
vanish independently gives the two equations
\begin{equation}
\label{keyeqs}
v_\alpha=\epsilon_{\alpha\beta\gamma}\lambda^\beta r^\gamma=\lambda\times r,~~~
\phi_\alpha=\epsilon_{\alpha\beta\gamma}\lambda^\gamma v^\beta=-\lambda\times v,
\end{equation}
respectively;  thus (\ref{begineqs}) and (\ref{keyeqs})
give all three possible cross products.
Transvection and substitution of (\ref{keyeqs}) gives
\begin{eqnarray}
\label{anseqs}
&&0=\lambda\cdot v=\lambda\cdot r=\lambda\cdot\phi=v\cdot r\phi_r,\\
&&\lambda^2=\frac{\phi_r}{r},~~~~~
v^2=\lambda^2r^2,~~~~~
\phi^2=\lambda^4r^2,\nonumber\\
&&\phi_\alpha=\lambda^2r_\alpha,~~~
\lambda_\alpha=\frac{L_\alpha}{\sum mr^2},~~~
[\lambda rv]\equiv\lambda\cdot(r\times v)=v^2,\nonumber
\end{eqnarray}
the first line of (\ref{anseqs}) gives all three possible dot products vanish;
all double cross products vanish.

The equation for $v$ in (\ref{keyeqs}) entails no motion along the $r$ or $\lambda$ axes
as $\lambda$ is proportional to $L$ (ix of \ref{anseqs}) and as $L$ is normal to the
ecliptic this means that motion is restricted to the ecliptic.
The equation $v\cdot r$ (iv of \ref{anseqs}) means that there is no radial change in $v$
so that the orbits are circular.

Some points arising are:
i)the stress tensor formed by metrical variation of (\ref{varl})
has a conservation equation that leads to no neat poisson equation;
ii)cannot use the equation for $\lambda_\alpha$ to replace $\lambda$ with $L$
in the lagrangian because of the $r$ dependence;
iii)no $\delta\lambda$ when varying the lagrangian;
iv)no $\delta x$,  as this just gives the differential of the $\delta v$ variation;
v)corresponding phase space lagrangian and
hamiltonian $v_\alpha-->p_\alpha/m$ gives nothing new,
in particular all poison brackets vanish;
vi)replacing P.E.=$\phi$ with scalar field $\sigma=2GM/\sqrt{\phi}$ leads to a ''mixed''
system in which the test particles and background gravitation are jumbled up,  so it becomes
unclear what aspect of the system one is studying the test particles or the background;
vii)seems that that is it,  cannot say anything more without a different lagrangian,
in particular cannot say anything about the time evolution of the system;
viii)exceptional cases are constant gravitational potential so that $\phi_r=0$,
in which case $v\cdot r$ does not necessarily vanish;
$v=0$ in which case there is no rotation then if the gravitational potential is repulsive
the particles move away from the original and if the gravitational potential is attractive
then the particles move in to the origin which could be thought of as dimensional reduction
to $d=0$.
\section{Four dimensional relativistic solar system description.}\label{fourd}
Now the standard picture (\ref{begineqs}) is generalized from newtonian $d=3$ space to $d=4$
K\"ottler (\ref{kottler}) spacetime,  which is Schwarzschild spacetime page \cite{ HE}
page 149 with cosmological constant.
The line element is
\begin{equation}
\label{kottler}
ds_4^2=-\left(1-\frac{2GM}{r}+\frac{\Lambda}{3}r^2\right)dt^2
+\frac{dr^2}{\left(1-\frac{2GM}{r}+\frac{\Lambda}{3}r^2\right)}+r^2d\Sigma_2^2,
\end{equation}
and the newtonian potential is
\begin{equation}
\label{wfpe}
g_{tt}=-1-2\phi,
\end{equation}
K\"ottler spacetime has the timelike Killing vector $T^a=\delta^a_t$
which is not of unit size but rather $T^aT_a=g_{tt}$.

To proceed generalize (\ref{varl}) term by term.
Firstly one has to choose a KE,  the KE
of (\ref{begineqs}) is retained but with $v$ now a four vector rather than a three vector,
this means that a lot of factors are assumed negligible such as:
i)lorentz-dirac,  deWitt-Brehme and Hobbs terms \cite{mdr12},
ii)dilation terms in $1-v^2/c^2$,
iii)whether or not the particle paths should be geodesic.
Secondly one has to choose a PE,  the newtonian potential (\ref{wfpe})
is used rather solutions to the d'Alembertian in this background as they do not allow separate
vanishing of $m~ \&~ \lambda$,
again there are a lot of more complicated expressions one could use.
Thirdly one has to generalize the angular momentum of the particles to four dimensions,
the simplest choice is
\begin{equation}
\label{kotL}
L_\alpha=m\epsilon_{\alpha\beta\gamma\delta}r^\beta v^\gamma T^\delta
=m\epsilon_{\alpha\beta\gamma t}r^\beta v^\gamma
\end{equation}
where $T$ is the timelike Killing vector of the spacetime.
Similarly
\begin{equation}
\label{kotkeyeqs}
v_\alpha=\epsilon_{\alpha\beta\gamma\delta}\lambda^\beta r^\gamma T^\delta,~~~
\phi_\alpha=\epsilon_{\alpha\beta\gamma\delta}\lambda^\gamma v^\beta T^\delta,
\end{equation}
Now the equations (\ref{anseqs}) follow as before,  in particular:
there is no correctional term from $T$ being of non-unit size as it is subscripted rather than
mixed each time,
there is no additional combinatorial term from the increase in dimension of the permutation
symbol as the fourth index is fixed at $t$ rather than being free to take all four dimensional
values.

\section{Five dimensional brane description}\label{fived}
Half of five dimensional de Sitter spacetime,  compare \cite{HE} page 125,  is
\begin{equation}
\label{dsxyzchi}
ds_5^2=\exp\left(-2\frac{\chi}{\alpha}\right)ds_4^2+d\chi^2,
\end{equation}
where $ds_4^2$ can be taken to be minkowski spacetime in rectilinear coordinates and
$\alpha^2=-4/\Lambda$.
The coordinate vectors
$\Xi^\alpha=\delta^\alpha_\chi,~T^\alpha=\delta^\alpha_t,~
X^\alpha=\delta^\alpha_x,Y^\alpha=\delta^\alpha_y,~Z^\alpha=\delta^\alpha_z$,
have non-unit size,  they are all killing vectors except $\Xi$ which is a conformal
killing vector,  they all have non-vanishing acceleration and shear.
The geometric configuration is different from the previous cases as instead of $r$ being
a distance to a mass at the origin one now has $\chi$ being the distance to some point in
the ''bulk''.
Again one first has to choose a KE,  again (\ref{begineqs}) is chosen except with $v$
now being a five vector, all possible correctional factors are taken to act only at
higher order.
For PE one again uses (\ref{wfpe}).

The generalization of the angular momenta $L$ (iv of \ref{begineqs})
is not so straightforward,  a general choice is
\begin{equation}
\label{ds5L}
L_\alpha^a
=m\epsilon_{\alpha\beta\gamma\delta\epsilon}\chi^\beta v^\gamma T^\delta X^a_\epsilon,
\end{equation}
where $a=x,y,z$,  thus $L$ is now a set of three 5-vectors for each particle $i$.
The equations generalizing (\ref{anseqs}) become complicated and give information
about choice of index $a$ rather than the problem in hand.
So we choose the simplest vector
\begin{equation}
\label{simp5L}
L^x_\alpha=\epsilon_{\alpha\beta\gamma tx}\chi^\beta v^\gamma,
\end{equation}
and variational term $-\lambda_a^\alpha\delta L^x_\alpha$.
These give (\ref{keyeqs},\ref{anseqs}) but with $r\leftrightarrow\chi$,
there are no new combinatorial factors.
In particular the fourth equality of (\ref{anseqs}) follows through showing that there
is no motion in the $\chi$ direction and this is the required result.
There are a number of problems assoicated with the choice
(\ref{ds5L}) and varied lagrangian term
$-\bar{g}_{ab}\lambda^b_\alpha\delta L^{a\alpha}$:
i)one should also vary with respect to
$\bar{g},~T,~X$ and assuming separate variations vanish give $L=0$ and the problem
degenerates;
ii)one could add compensating fields however then the problem no longer
remains the simplest way of proceeding;  also with a free-indexed five dimensional
permutation symbol the problem becomes computationally complex;
iii)the first three or transvection equations of (\ref{anseqs})
do not hold unless $\lambda^{\alpha[a}\lambda^{|\beta|b]}=0$ the substitution equation
become complicated.
\section{Conclusion}\label{conc}
Test matter with angular momenta (\ref{simp5L}) rotating around a fifth dimension is confined
to four dimensional spacetime.
This can be thought of as providing a qualitative description of why matter is
restricted to four dimensions and also why non-gravitational forces act only in four dimensions.
The angular momenta (\ref{simp5L}) involve the permutation symbol as do chern-simons terms
but there is no immediate relationship as chern-simons terms involve partial derivatives.
Although the above provides a qualitative description it does not provide any quantitative
relationship such as between the cosmological constant and the size of the angular momenta
(\ref{simp5L}).
\section{Acknowledgements}\label{ack}
The idea for this paper occurred while doing an edx anu mooc.
I would like to thank prof.Tremaine for sending me a copy of a slide describing the solar
system case and prof.Bombelli for discussion of exceptional cases.

\end{document}